\documentclass{Interspeech}

\interspeechcameraready 

\title{VoiceMark: Zero-Shot Voice Cloning-Resistant Watermarking Approach Leveraging Speaker-Specific Latents}

\author[affiliation={1,2}]{Haiyun}{Li}
\author[affiliation={1,2,*}]{Zhiyong}{Wu}
\author[affiliation={3}]{Xiaofeng}{Xie}
\author[affiliation={1}]{Jingran}{Xie}
\author[affiliation={1}]{Yaoxun}{Xu}
\author[affiliation={2,*}]{Hanyang}{Peng}

\affiliation{Shenzhen International Graduate School}{Tsinghua University}{China}
\affiliation{}{Pengcheng Laboratory}{China}
\affiliation{}{Independent Researcher}{China}
\email{lihaiyun24@mails.tsinghua.edu.cn, zywu@sz.tsinghua.edu.cn, xiexiaofeng1926@gmail.com, \{xjr21, xuyx22\}@mails.tsinghua.edu.cn, penghy@pcl.ac.cn}


\usepackage{comment}
\usepackage{graphicx}
\usepackage{subfig}
\usepackage{hyperref}
\begin{document}
\maketitle
\renewcommand{\thefootnote}{\fnsymbol{footnote}}
\footnotetext[1]{Corresponding authors.}

% the abstract here must exactly match the abstract entered into the paper submission system
\begin{abstract}
Voice cloning (VC)-resistant watermarking is an emerging technique for tracing and preventing unauthorized cloning. Existing methods effectively trace traditional VC models by training them on watermarked audio but fail in zero-shot VC scenarios, where models synthesize audio from an audio prompt without training. 
To address this, we propose VoiceMark, the first zero-shot VC-resistant watermarking method that leverages speaker-specific latents as the watermark carrier, allowing the watermark to transfer through the zero-shot VC process into the synthesized audio. Additionally, we introduce VC-simulated augmentations and VAD-based loss to enhance robustness against distortions.
Experiments on multiple zero-shot VC models demonstrate that VoiceMark achieves over 95\% accuracy in watermark detection after zero-shot VC synthesis, significantly outperforming existing methods, which only reach around 50\%. See our code and demos at: \href{https://huggingface.co/spaces/haiyunli/VoiceMark}{https://huggingface.co/spaces/haiyunli/VoiceMark}.
\end{abstract}

\section{Introduction}
\begin{figure}[ht]
\centering
\includegraphics[width=1.0\linewidth]{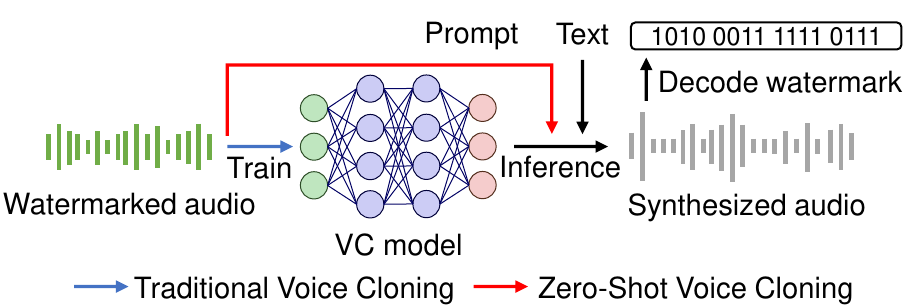}
\caption{Comparison of data flow for embedding watermarks in traditional voice cloning and zero-shot voice cloning.}
\label{fig:vc}
\end{figure}

Voice cloning (VC)-resistant watermarking is an emerging technique for tracing and preventing unauthorized cloning. Artists can embed such watermarks into their copyrighted recordings, ensuring that even if their voice is cloned into new audio, the watermark remains intact, thereby tracing and preventing unauthorized cloning. Recent research \cite{timbrewatermarking-ndss2024} has explored this application, confirming that most traditional VC models trained on watermarked audio will synthesize audio that retains the watermark. However, with the rapid development of zero-shot VC models such as CosyVoice \cite{du2024cosyvoice}, F5-TTS \cite{chen2024f5}, and MaskGCT \cite{wang2024maskgct}, highly realistic cloned audio can now be synthesized without requiring training, using only a few seconds of audio prompt. In this new scenario, we cannot embed watermarks into VC models through training, as the watermarked audio is directly used for inference, as illustrated in Figure \ref{fig:vc}. This means a zero-shot VC-resistant watermark is needed — one that can be directly transferred to the synthesized audio during zero-shot VC inference using a single watermarked audio prompt. Currently, zero-shot VC-resistant watermarking has not yet been studied.

Traditional audio watermarking has been developed over many years, with techniques such as spread spectrum \cite{Bender_Gruhl_Morimoto_Lu_1996} and echo hiding \cite{Gruhl_Lu_Bender_1996}. In recent years, more advanced deep learning-based approaches have been proposed \cite{chen2023wavmark, san2024proactive}, demonstrating greater robustness compared to traditional methods. However, these watermarking methods are mainly designed to resist traditional audio editing, such as noise addition, filtering, and compression, with limited focus on resistance to VC models. Some methods \cite{timbrewatermarking-ndss2024, li2024proactive} propose VC-resistant watermarking techniques designed to resist traditional VC models. They validates that training traditional VC models \cite{kim2021conditional,renfastspeech,shen2018natural} or commercial tools \cite{zhang2022paddlespeech,VoiceCloningApp} on watermarked audio enables robust watermark detection from the resulting synthesized audio. However, in the context of zero-shot VC-resistant watermarking, such methods face several limitations:

\begin{itemize}
\item \textbf{Inability to embed watermarks via training}: Zero-shot VC models eliminate the possibility of embedding watermarks into model parameters through training, causing existing VC-resistant watermarking methods to fail.
\item \textbf{Failure to retain watermarks during synthesis}: Audio synthesized by zero-shot VC models exhibits significant differences from the prompt, including changes in content, length, and speed. These transformations disrupt or filter out the watermark, making it undetectable in the synthesized audio.
\end{itemize}
These limitations hinder the direct application of existing watermarking methods for traceability in zero-shot VC, exposing individuals' voices to the risk of unauthorized cloning.

\begin{figure*}[ht]
\centering
\includegraphics[width=0.89\textwidth]{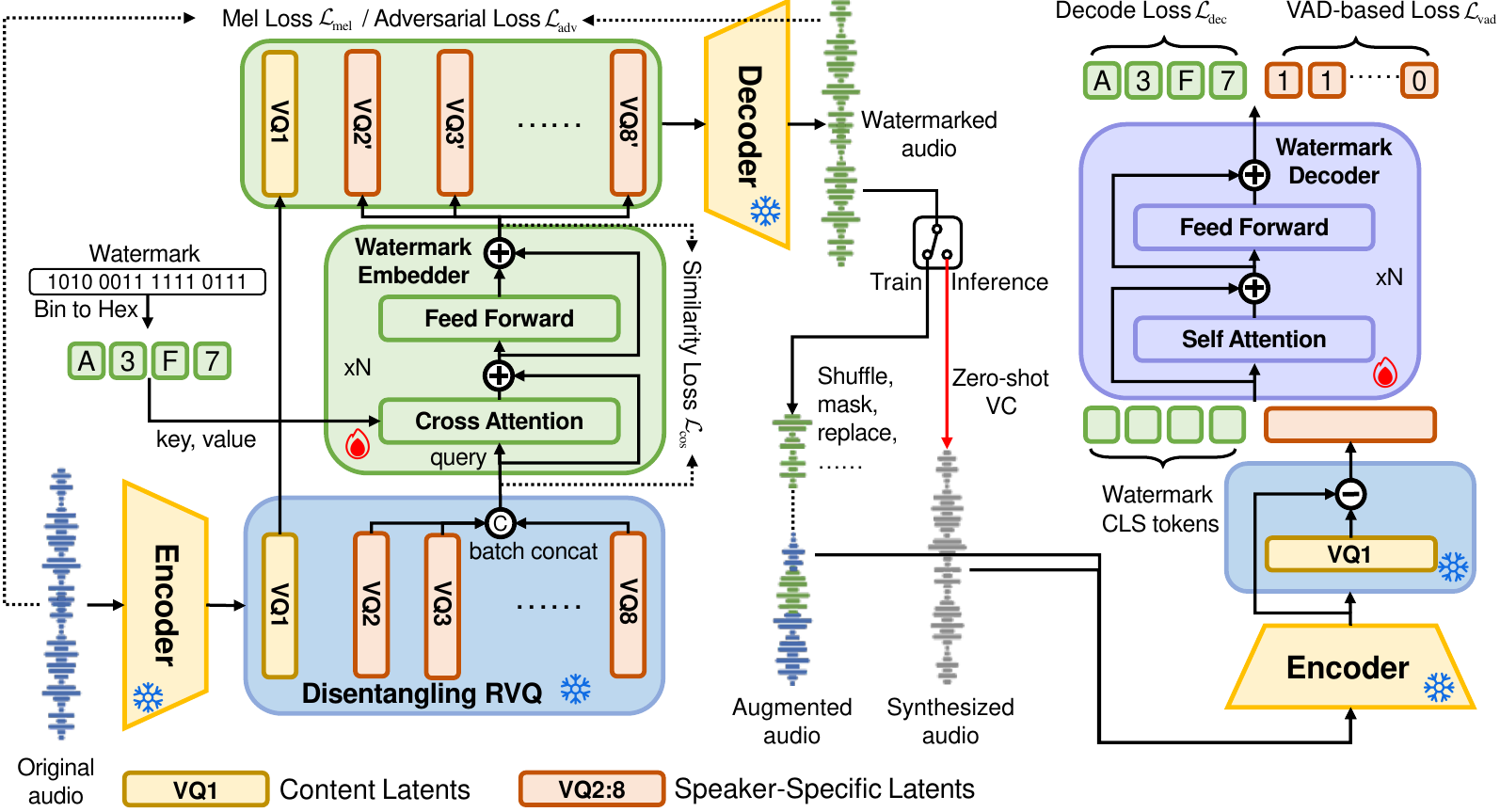}
\caption{The overall architecture of our proposed VoiceMark.}

\label{fig:architecture}
\end{figure*}

To mitigate these risks, we aim to design a robust zero-shot VC-resistant watermarking method that does not rely on training VC models yet enables traceability for zero-shot VC. We observe that in zero-shot VC, to clone a speaker's voice, models must extract speaker-specific information (timbre, pitch, prosody, etc.) while discarding content from the audio prompt. This process typically occurs in the latent space compressed by neural codec-based methods \cite{defossez2022high,junaturalspeech,zhang2023speechtokenizer} where the model implicitly or explicitly disentangles the speaker-specific latents to synthesize cloned audio. This implies that for higher speaker similarity, the model must consistently transfer speaker-specific latents from the original audio prompt to the synthesized audio.

Therefore, the key idea of our work is to leverage speaker-specific latents as the watermark carrier by adopting a neural codec model to disentangle them, embed the watermark, and reconstruct watermarked audio. The watermark is then transferred along with the speaker-specific latents to the synthesized audio during the zero-shot VC inference, thereby enabling robust zero-shot VC-resistant watermarking. To achieve this goal, we must address two primary challenges:
1) Speaker-specific latents span only speech-containing frames, while a large portion of the audio consists of silence and voiceless frames, which lack such latents. This makes conventional frame-by-frame embedding methods ineffective in leveraging speaker-specific latents.
2) The VC process alters content, duration, and speed while also distorting speaker-specific latents, causing the watermark to remain in only certain frames and potentially be incomplete. This makes accurate watermark detection challenging.

In this paper, we propose VoiceMark, the first zero-shot VC-resistant watermarking method that significantly improves resistance across multiple zero-shot VC models compared to state-of-the-art (SOTA) audio watermarking methods. Our main contributions are summarized as follows:
\begin{itemize}

\item VoiceMark introduces a watermark embedder model based on speaker-specific latents. It disentangles speaker-specific latents using a pre-trained residual vector quantization (RVQ) model, embed the watermark into these latents, and generates watermarked audio. We design a voice activity detection (VAD)-based loss function to guide the model in identifying frames containing speaker-specific latents and adaptively embedding the watermark through cross-attention mechanisms. 
\item VoiceMark proposes a robust watermark decoder model. To simulate the potential distortions of the zero-shot VC synthesis on the watermark, its training incorporates augmentation techniques, such as masking, shuffling, replacing, and neural encoding/decoding. Using a global transformer, the decoder recovers the watermark from the speaker-specific latents of the entire audio to enable robust detection.
\item We conduct extensive experiments on multiple zero-shot VC models and traditional audio editing methods to evaluate the effectiveness of VoiceMark. Our results demonstrate that VoiceMark achieves over 95\% accuracy in watermark detection after zero-shot VC synthesis, which significantly outperforms existing watermarking methods that reach around 50\%.

\end{itemize}

\section{Methodology}

VoiceMark consists of three main components: an encoder-decoder RVQ model, a cross-attention-based watermark embedder, and a transformer-based watermark decoder, as shown in Figure \ref{fig:architecture}. The RVQ model disentangles speaker-specific latents and reconstructs watermarked audio. The embedder embeds the watermark into speaker-specific latents. The decoder extracts the watermark from latents of the synthesized audio.

\subsection{Disentangling Speaker-Specific Latents} 

The RVQ model, inspired by SpeechTokenizer~\cite{zhang2023speechtokenizer}, leverages HuBERT \cite{hsu2021hubert} latents as a semantic teacher to distill content latents into the first VQ layer, thereby disentangling speaker-specific latents into the remaining layers (VQ 2 to 8).

Given an input audio $x$, the encoder compresses it into a latent sequence $l = E(x)$, where $l \in \mathbb{R}^{t \times d}$, $t$ is the number of frames, and $d$ is the latent dimension. The sequence is then quantized by an 8-layer RVQ model, producing a set of quantized latents $\{z_1, z_2, \dots, z_8\}$, where $z_1$ represents content latents and $\{z_2, \dots, z_8\}$ correspond to speaker-specific latents. The content latents $z_1$ remain unmodified, while the speaker-specific latents are used for subsequent watermark embedding.

\subsection{Watermark Embedding}
Given the speaker-specific latents $\{z_2, \dots, z_8\}$ and an $n$-bit watermark, the watermark is first converted into a hexadecimal sequence $w \in \{0, 1, \dots, 15\}^{n / 4}$ to reduce its length. The sequence $w$ is then projected into the latent space of dimension $d$, resulting in $w' \in \mathbb{R}^{(n / 4) \times d}$. The speaker-specific latents ${z_2, \dots, z_8}$ and $w'$ are then fed into a 4-layer cross-attention embedder $W_e(\cdot, \cdot)$ based on the transformer decoder \cite{vaswani2017attention}, which produces the watermarked speaker-specific latents:
\begin{equation} 
    \{\hat{z}_2, \dots, \hat{z}_8\} = W_e(\{z_2, \dots, z_8\}, w').
\end{equation} 

Here, $\{z_2, \dots, z_8\}$ serve as queries, while $w'$ serves as keys and values. 
After embedding the watermark, all latents are aggregated as $\hat{z} = z_1 +  \sum_{i=2}^8 \hat{z}_i$, where $\hat{z}$ represents the watermarked latents. These are then passed into the decoder $D(\hat{z})$ to reconstruct the watermarked audio $\hat{x}$.

\subsection{VC-Simulated Augmentation}  
The zero-shot VC synthesis process can significantly distort watermarked audio. To enhance the model’s robustness, we introduce several augmentations that simulate these distortions:  
\begin{enumerate}
    \item \textbf{No watermark in silent frames}. We mask certain frames to zero with a probability of 20\%.
    \item \textbf{Completely different content}. We randomly shuffle the audio in 50 ms windows with a 50\% probability.
    \item \textbf{Partial watermark filtering}. We replace 50 ms segments with the original audio at a probability of 50\%.
    \item \textbf{Neural codec encoding and decoding}. We encode and decode the audio using EnCodec~\cite{defossez2022high}.
    \item \textbf{Audio perturbation}. We apply speed perturbation, amplitude scaling, filtering, or resampling with 10\% probability.

\end{enumerate}  
This process generates augmented audio $\tilde{x}$ to train the model and improve its robustness against distortions.

\subsection{Watermark Decoding}  

The watermark decoder, denoted as $W_d(\cdot)$, is a transformer model for extracting the watermark from speaker-specific latents. During inference, it processes audio synthesized by zero-shot VC models, while during training, it learns from augmented audio $\tilde{x}$ without exposure to any VC models. The input audio is first compressed into latents $l$. The content latents from the first VQ layer ($z_1$) are subtracted from $l$ to obtain the speaker-specific latents, denoted as $l_s = l - z_1$. 

For watermark decoding, we use learnable CLS tokens, denoted as $c \in \mathbb{R}^{(n/4) \times d}$, corresponding to the $n / 4$-length hexadecimal watermark. These tokens are concatenated with $l_s$ and fed into $W_d(\cdot)$, which employs self-attention to extract watermark information from $l_s$:
\begin{equation}
    (\hat{w}, \hat{p}) = W_d(c \oplus l_s),
\end{equation}
where $\oplus$ denotes the concatenation operation, $\hat{w} \in \mathbb{R}^{(n/4) \times 16}$ is the decoded hexadecimal watermark, with each row representing a softmax probability distribution over 16 categories, $\hat{p} \in \mathbb{R}^{t}$ is the sigmoid probability of a frame containing both speech and a watermark ($\hat{p} = 1$) or neither ($\hat{p} = 0$). The decoded hexadecimal watermark is first processed with argmax, then converted into an $n$-bit binary sequence.

\subsection{Training Loss}  
We design multiple loss functions to ensure accurate watermark detection while preserving audio quality:

\textbf{VAD-based Loss.} We use a dual-threshold VAD method~\cite{rabiner1978digital} to compute binary cross entropy $\mathcal{L}_{\text{vad}}$ on $\hat{p}$. Frames containing both speech and a watermark are labeled as 1, while silent, masked or replaced frames are labeled as 0. This guides the embedder and decoder to focus on watermarked speech.

\textbf{Quality Loss.} We preserve speaker consistency with a cosine similarity loss $\mathcal{L}_{\text{cos}}$ between speaker-specific latents before and after watermark embedding. For perceptual quality, we apply a multi-scale Mel spectrogram loss $\mathcal{L}_{\text{mel}}$ and an adversarial loss $\mathcal{L}_{\text{adv}}$ to refine the reconstructed audio $\hat{x}$ \cite{defossez2022high}.

\textbf{Decoding Loss.} The watermark decoding loss $\mathcal{L}_{\text{dec}}$ is computed using the cross-entropy loss on $\hat{w}$, ensuring accurate recovery of the hexadecimal watermark.

The total loss is a weighted sum of these components:
\begin{equation}
    \mathcal{L} = \lambda_{\text{vad}} \mathcal{L}_{\text{vad}}+ \lambda_{\text{cos}} \mathcal{L}_{\text{cos}} + \lambda_{\text{mel}} \mathcal{L}_{\text{mel}} + \lambda_{\text{adv}} \mathcal{L}_{\text{adv}} + \lambda_{\text{dec}} \mathcal{L}_{\text{dec}},
\end{equation}
where $\lambda_{\text{vad}}, \lambda_{\text{cos}}, \lambda_{\text{mel}}, \lambda_{\text{adv}}, \lambda_{\text{dec}}$ are hyperparameters.

\begin{table*}[ht]
\centering
\caption{Performance Evaluation on Zero-Shot VC Models and Traditional Editing.}
\resizebox{\textwidth}{!}{%
\begin{tabular}{lccccccccc}
\toprule
 & \multicolumn{3}{c}{Zero-Shot VC Models (ACC ↑ / FAR ↓)} & \multicolumn{6}{c}{Traditional Editing (ACC ↑ / FAR ↓)} \\
\cmidrule(lr){2-4} \cmidrule(lr){5-10}
Method & CosyVoice \cite{du2024cosyvoice} & F5-TTS \cite{chen2024f5} & MaskGCT \cite{wang2024maskgct} & EnCodec \cite{defossez2022high} & Resample & Amplitude & Filter & White & MP3 \\
\midrule
AudioSeal (2024) \cite{san2024proactive} & 0.508/0.979 & 0.513/0.977 & 0.506/0.973 & 0.936/0.052 & 1.000/0.000 & 1.000/0.000 & 1.000/0.000 & 1.000/0.000 & 1.000/0.000 \\
WavMark (2023) \cite{chen2023wavmark}  & 0.499/1.000 & 0.499/1.000 & 0.499/1.000 & 0.498/1.000 & 1.000/0.000 & 1.000/0.000 & 0.711/0.289 & 0.938/0.058 & 1.000/0.000 \\
Timbre (2024) \cite{timbrewatermarking-ndss2024} & 0.499/0.981 & 0.539/0.954 & 0.527/0.966 & 0.696/0.726 & 1.000/0.000 & 1.000/0.000 & 0.989/0.022 & 1.000/0.000 & 1.000/0.000 \\
VoiceMark (Ours)    & \textbf{0.964}/\textbf{0.112} & \textbf{0.979}/\textbf{0.070} & \textbf{0.957}/\textbf{0.141} & \textbf{0.985}/\textbf{0.044} & 0.988/0.049 & 0.995/0.014 & 0.987/0.036 & 0.965/0.109 & 0.973/0.102 \\
\bottomrule
\end{tabular}
}
\label{tab:performance}
\end{table*}

\section{Experiments}

\subsection{Implementation Details}
For the RVQ model, we use the pretrained SpeechTokenizer\footnote{https://huggingface.co/fnlp/SpeechTokenizer}. $W_e$ employs a 4-layer, 1-head, 256-dimensional Transformer Decoder, while $W_d$ adopts an 8-layer, 1-head, 512-dimensional Transformer Encoder \cite{vaswani2017attention}. The watermark bit length is set to 16. Hyperparameters are set as: $\lambda_{\text{vad}}=1$, $\lambda_{\text{cos}}=2$, $\lambda_{\text{mel}}=2$, $\lambda_{\text{adv}}=1$, and $\lambda_{\text{dec}}=1$, with larger weights for $\lambda_{\text{cos}}$ and $\lambda_{\text{mel}}$ to preserve audio quality. The model is trained for 30 epochs using Adam \cite{kingma2014adam} optimizer with a learning rate of $5e^{-5}$.

\subsection{Baselines}
We compare VoiceMark with three SOTA watermarking methods: AudioSeal \cite{san2024proactive}, WavMark \cite{chen2023wavmark}, and Timbre \cite{timbrewatermarking-ndss2024}. The related method \cite{li2024proactive} is not included since the code is not public.

\subsection{Datasets}
We use the VCTK \cite{Yamagishi2012} (train and test) and Librispeech \cite{panayotov2015librispeech} (test) datasets. The test set consists of 2,000 unseen VCTK samples and 2,600 Librispeech samples, totaling 4,600 samples from 150 speakers.

\subsection{Metrics}
We evaluate watermark detection using two metrics: bitwise accuracy (ACC) and False Attribution Rate (FAR). ACC is the ratio of correctly decoded bits to the total number of bits. FAR simulates real-world multi-candidate identification by comparing the Hamming distance of each decoded watermark to 100 candidates (1 ground truth and 99 random). A false attribution occurs if the closest match is not the ground truth.

For audio quality, we assess objective metrics including perceptual evaluation of speech quality (PESQ) \cite{rix2001perceptual}, scale-invariant signal-to-noise ratio (SI-SNR), and short-time objective intelligibility (STOI) \cite{taal2010short}. Subjective evaluation is conducted using similarity mean opinion score (SMOS), where 15 subjects rate 40 samples on a 1-5 scale. A SMOS of 4 or higher indicates high similarity to the original audio.

\subsection{Performance Evaluation}
Table \ref{tab:performance} compares VoiceMark’s watermark detection performance with other SOTA methods under zero-shot VC models (CosyVoice \cite{du2024cosyvoice}, F5-TTS \cite{chen2024f5}, and MaskGCT \cite{wang2024maskgct}) and traditional audio editing. The text prompts for zero-shot VC are randomly selected from the test set.

For zero-shot VC models, VoiceMark consistently outperforms all methods. Other approaches exhibit an ACC close to 0.5 and an FAR near 1.0, indicating near-random decoding in zero-shot VC scenarios. This demonstrates that existing methods fail to retain the watermark in zero-shot VC, whereas VoiceMark ensures robust watermark traceability.

For traditional audio editing, we use AudioCraft’s implementation\footnote{https://github.com/facebookresearch/audiocraft} with default parameters, where Amplitude is the average of boost and duck, and Filter is the average of band, high, and low-pass filters. The results show that VoiceMark performs comparably to other methods and outperforms them after EnCodec processing. We observe that VoiceMark doesn't achieve 1.0 ACC, likely due to our training solely on the VCTK dataset. VCTK has many speakers suitable for VC-related tasks, but it is a relatively small, clean dataset, and the distorted audio from editing may degrade our performance.

\begin{table}[ht]
\small
\centering
\caption{Ablation Study}
\setlength{\tabcolsep}{16pt}
\begin{tabular}{lcc}
\toprule
Method & ACC ↑ & FAR ↓ \\
\midrule
AS-Emb + VM-Dec & 0.663 & 0.906 \\
VoiceMark & \textbf{0.964} & \textbf{0.112} \\
\hspace{10pt} w/o VAD-base Loss & 0.478 & 0.980 \\
\hspace{10pt} w/o Augmentation & 0.626 & 0.924 \\
\bottomrule
\end{tabular}
\label{tab:ablation_study}
\end{table}

\subsection{Ablation Study}
VoiceMark incorporates three key innovations: speaker-specific latents watermarking, VC-simulated augmentation, and VAD-based loss. We conduct ablation studies on CosyVoice \cite{du2024cosyvoice} to validate their necessity, as shown in Table \ref{tab:ablation_study}.

To assess the impact of speaker-specific latents watermarking, we keep other modules unchanged but replace the watermark embedder with the AudioSeal embedder \cite{san2024proactive} (AS-Emb + VM-Dec), which is based on a generic architecture, where the synthesized watermark is directly added to the original waveform. Results show that our latent-based design significantly enhances zero-shot VC resistance. Additionally, removing VC-simulated augmentation and VAD-based loss causes a performance drop, confirming the necessity for robust watermarking.

\subsection{Audio Quality Assessment}
Table \ref{tab:audio_quality} presents the audio quality evaluation results for different methods. AudioSeal directly adds the generated watermark to the original audio waveform, while EnCodec and SpeechTokenizer are neural codec models that reconstruct the audio from a latent space. VoiceMark, leveraging a neural codec architecture for watermarking, integrates the watermarking process within the codec framework. 

The results show that while VoiceMark's audio quality is lower than AudioSeal's, it performs comparably to SpeechTokenizer and significantly outperforms EnCodec.

\begin{table}[ht]
\centering
\caption{Audio Quality Assessment. 
(W): General watermarking directly added to original waveforms.
(N): Neural codec models.
(W-N): Watermarking using neural codec architecture.}
\resizebox{\linewidth}{!}{%
\begin{tabular}{lcccc}
\toprule
Method & PESQ ↑ & SI-SNR ↑ & STOI ↑ & SMOS ↑ \\
\midrule
AudioSeal (W) \cite{san2024proactive} & 4.32 & 26.69 & 0.99 & 4.67$\pm$0.10 \\
\midrule
EnCodec (N) \cite{defossez2022high} & 1.62 & -0.62 & 0.80 & 2.21$\pm$0.15 \\
SpeechTokenizer (N) \cite{zhang2023speechtokenizer} & 2.58 & 1.64 & 0.89 & 4.63$\pm$0.10 \\
VoiceMark (W-N) & 2.20 & 2.01 & 0.89 & 4.25$\pm$0.13 \\
\bottomrule
\end{tabular}
}
\label{tab:audio_quality}
\end{table}

\begin{figure}[h]
    \centering
    \subfloat[Original]{\includegraphics[width=0.33\linewidth]{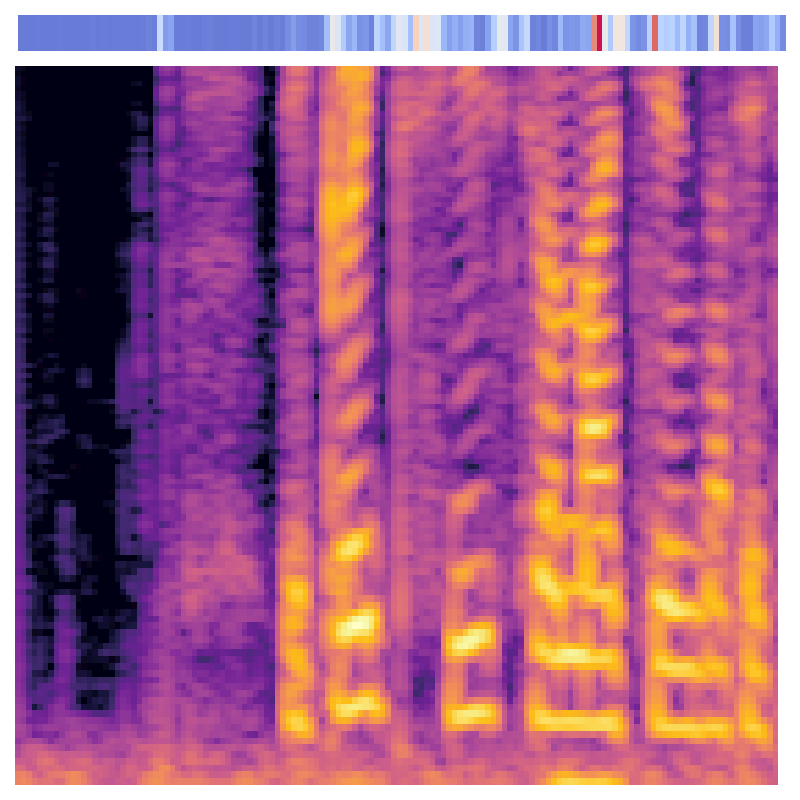}} 
    \subfloat[AudioSeal \cite{san2024proactive}]{\includegraphics[width=0.33\linewidth]{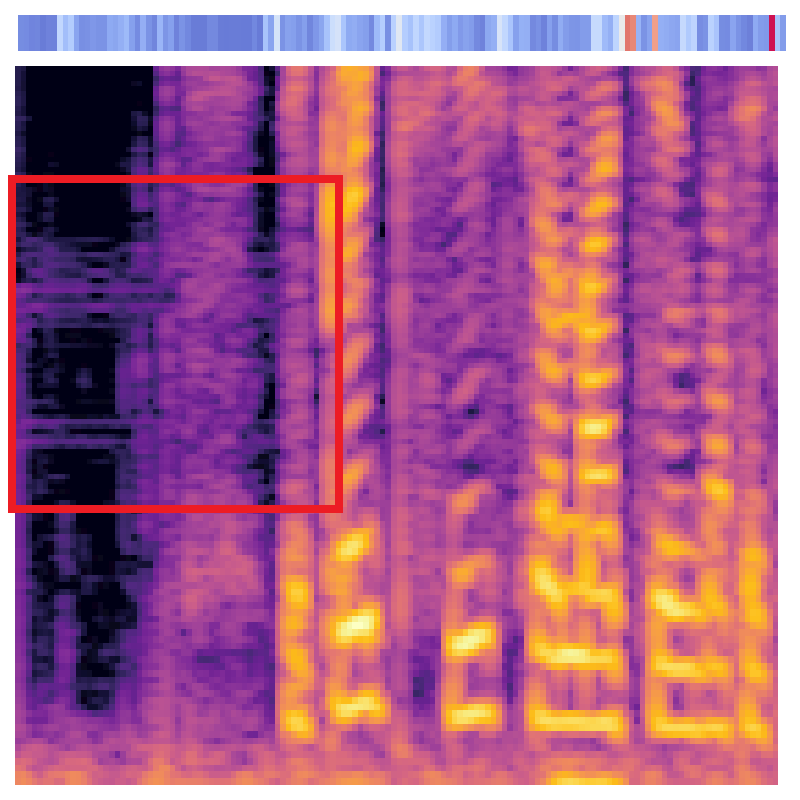}} 
    \subfloat[WavMark \cite{chen2023wavmark}]{\includegraphics[width=0.33\linewidth]{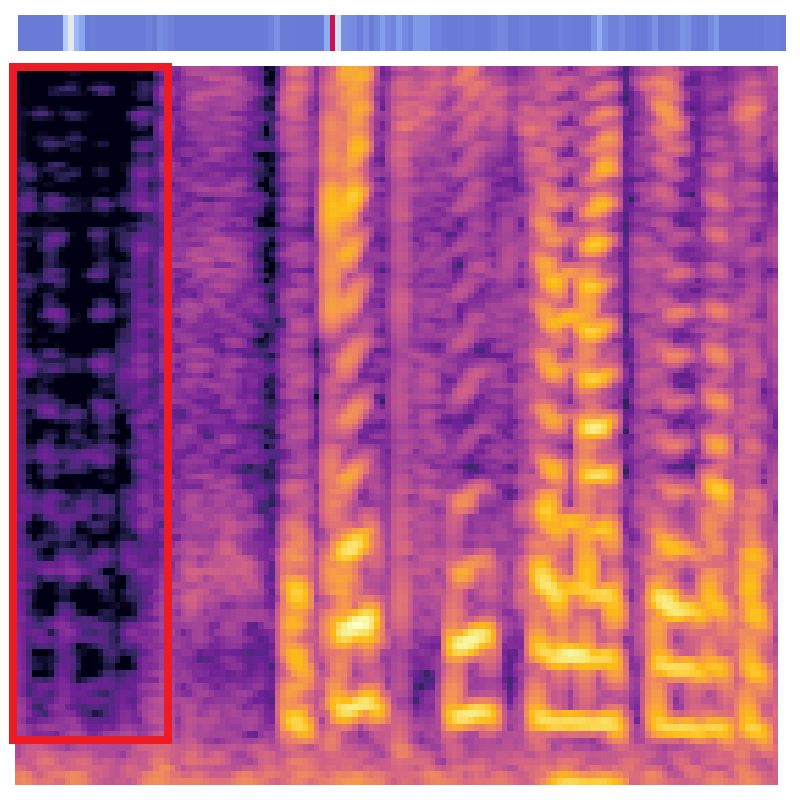}} \\
    \subfloat[Timbre \cite{timbrewatermarking-ndss2024}]{\includegraphics[width=0.33\linewidth]{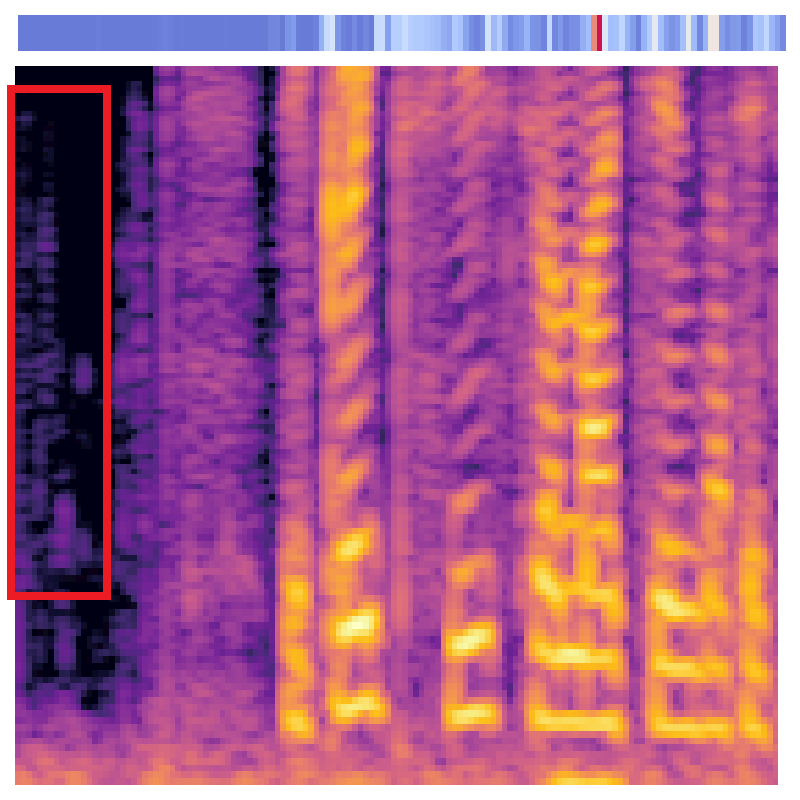}} 
    \subfloat[VoiceMark (Ours)]{\includegraphics[width=0.33\linewidth]{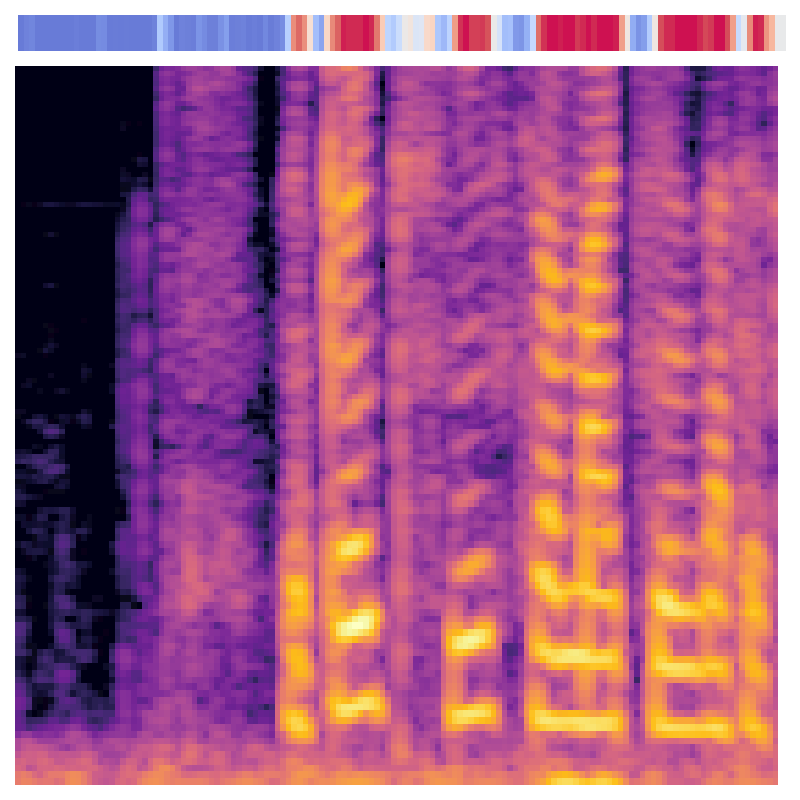}}
    \caption{Visualization of Mel Spectrograms.}
    \label{case_study}
\end{figure}

\subsection{Case Study}
To directly observe the impact of watermarking on audio, we visualize the mel spectrograms of different watermarking methods, as shown in Figure \ref{case_study}.

VoiceMark embeds the watermark within speaker-specific latents, subtly altering harmonics and formants, making it more difficult for attackers to detect. In contrast, spectrograms from other methods exhibit visible watermarking artifacts in certain frequency bands (highlighted in red boxes), which attackers can exploit to detect or remove the watermark.

In addition, we visualize watermark probability detected by VoiceMark as a top color band, with red for high and blue for low probability. In VoiceMark’s watermarked audio, the detected probability aligns precisely with speech segments, while no watermark appears in other samples, confirming its effective embedding within speech features.

\section{Conclusion}

In this work, we propose VoiceMark, the first zero-shot VC-resistant watermarking method that embeds watermarks into speaker-specific latents to achieve resistance to zero-shot VC models. Additionally, we incorporate VC-simulated augmentations and VAD-based loss to further enhance the robustness of VoiceMark. Experimental results show that VoiceMark significantly outperforms SOTA watermarking methods in retaining watermarks across multiple zero-shot VC models. 

\section{Acknowledgements}
This work is supported by National Natural Science Foundation of China (62076144).

\bibliographystyle{IEEEtran}
\bibliography{mybib}

\end{document}